\documentstyle[12pt]{article}
\def\Re{{\cal R \mskip-4mu \lower.1ex \hbox{\it e}\,}}
\def\Im{{\cal I \mskip-5mu \lower.1ex \hbox{\it m}\,}}

\def\etal{{\it et al.}}

\def\sub#1{_{\lower.25ex\hbox{$\scriptstyle#1$}}}
\def\sul#1{_{\kern-.1em#1}}
\def\sll#1{_{\kern-.2em#1}}
\def\sbl#1{_{\kern-.1em\lower.25ex\hbox{$\scriptstyle#1$}}}
\def\ssb#1{_{\lower.25ex\hbox{$\scriptscriptstyle#1$}}}
\def\sbb#1{_{\lower.4ex\hbox{$\scriptstyle#1$}}}

\def\to{\rightarrow}
\def\mh{\ifmmode m\sbl H \else $m\sbl H$\fi}
\def\mch{\ifmmode m_{H^\pm} \else $m_{H^\pm}$\fi}
\def\mt{\ifmmode m_t\else $m_t$\fi}
\def\mc{\ifmmode m_c\else $m_c$\fi}
\def\mz{\ifmmode M_Z\else $M_Z$\fi}
\def\mw{\ifmmode M_W\else $M_W$\fi}
\def\mws{\ifmmode M_W^2 \else $M_W^2$\fi}
\def\mhs{\ifmmode m_H^2 \else $m_H^2$\fi}
\def\mzs{\ifmmode M_Z^2 \else $M_Z^2$\fi}
\def\mts{\ifmmode m_t^2 \else $m_t^2$\fi}
\def\mcs{\ifmmode m_c^2 \else $m_c^2$\fi}
\def\mchs{\ifmmode m_{H^\pm}^2 \else $m_{H^\pm}^2$\fi}
\def\ztwo{\ifmmode Z_2\else $Z_2$\fi}
\def\zone{\ifmmode Z_1\else $Z_1$\fi}
\def\mtwo{\ifmmode M_2\else $M_2$\fi}
\def\mone{\ifmmode M_1\else $M_1$\fi}
\def\tb{\ifmmode \tan\beta \else $\tan\beta$\fi}
\def\xw{\ifmmode x\sub w\else $x\sub w$\fi}
\def\ch{\ifmmode H^\pm \else $H^\pm$\fi}
\def\lum{\ifmmode {\cal L}\else ${\cal L}$\fi}
\def\inpb{\ifmmode {\rm pb}^{-1}\else ${\rm pb}^{-1}$\fi}
\def\infb{\ifmmode {\rm fb}^{-1}\else ${\rm fb}^{-1}$\fi}
\def\epem{\ifmmode e^+e^-\else $e^+e^-$\fi}
\def\ppb{\ifmmode \bar pp\else $\bar pp$\fi}

\newskip\zatskip \zatskip=0pt plus0pt minus0pt
\def\matth{\mathsurround=0pt}

\def\atversim#1#2{\lower0.7ex\vbox{\baselineskip\zatskip\lineskip\zatskip
  \lineskiplimit 0pt\ialign{$\matth#1\hfil##\hfil$\crcr#2\crcr\sim\crcr}}}

\renewcommand{\thefootnote}{\fnsymbol{footnote}}

\begin{document} \begin{titlepage}
\setcounter{page}{1}
\thispagestyle{empty}
\rightline{\vbox{\halign{&#\hfil\cr
&ANL-HEP-PR-93-8\cr
&January 1993\cr}}}
\vspace{1in}
\begin{center}

{\Large\bf
Constraints on New Physics From Tevatron Dijet Data}
\footnote{Research supported by the
U.S. Department of
Energy, Division of High Energy Physics, Contract W-31-109-ENG-38.}
\medskip

\normalsize THOMAS G. RIZZO
\\ \smallskip
High Energy Physics Division\\Argonne National
Laboratory\\Argonne, IL 60439\\

\end{center}

\begin{abstract}

New results from the CDF Collaboration on the invariant mass distribution
of dijet events
observed at the Tevatron are used to constrain the masses and couplings of
new gauge bosons ($W'$ and $Z'$) as well as fundamental diquarks which can
occur in some extended electroweak models. In the case of new gauge bosons,
these new bounds are then compared to existing limits
which arise from searches for the leptonic decay modes of these particles
at the Tevatron.

\end{abstract}

\renewcommand{\thefootnote}{\arabic{footnote}} \end{titlepage}


Although the Standard Model(SM) has enjoyed enormous success in explaining
present data{\cite {smrev}}, it is generally believed that it cannot be the
whole story as it leaves too many unanswered questions. Models which go
beyond the standard scenario trying to address some of these issues
usually predict the existence of new degrees of
freedom, as yet unobserved, in the mass range not far above that of the
conventional $W$ and $Z$. Some of these states may eventually be produced
at existing or planned colliders.

One of the most common classes of such extended models predicts an
enlargement of the SM
gauge symmetry and thus new $Z'$ and/or $W'$ gauge bosons. These have been
searched for, with so far negative results, both directly at hadron colliders
via their leptonic decay
modes{\cite {ual,tevl}} and indirectly at LEP{\cite {LEP}} through the effects
of mixing. Since new gauge bosons generally have substantial couplings to
both quarks and
leptons{\cite {lphil}}, searches in the $Z',W'\to 2~jets$ channel also need
to be carried out as these are complimentary to the more conventional ones
involving only final state leptons, $e or \mu$.
Searches of this kind are, of course, more difficult due to the
presence of large QCD backgrounds, but these are not impossible to
overcome as has been shown by the UA2
Collaboration in the case of ordinary $W$ and $Z$ production{\cite {UA2}}.

Recently, the CDF Collaboration has made a precise measurement of the dijet
invariant mass spectrum at the Tevatron{\cite {tevjj}} and performed a
comparison with the predictions of QCD at
Next-to-Leading-Log(NLL) order for various structure function choices. From
this analysis they were able to place limits on the production of `narrow'
dijet resonances over a broad mass range. (Here, by `narrow' we mean that the
width to mass ratio, $\Gamma/M$, of a resonance is less than 0.1.) To display
the power of these measurements, they then showed that axigluons{\cite {axi}},
an octet of strongly coupled massive gauge bosons which are predicted in
a chiral version of QCD,
were excluded over a large range of hypothetical masses. Of course, other
particles which can couple to pairs of quarks and/or gluons which sufficient
strength might now also be excluded by this same set of data.

The purpose of the present work is to examine what additional constraints,
if any, this new Tevatron dijet data places on the existence of new gauge
bosons, $Z'$ and $W'$,
and fundamental diquarks (which can appear, for example, in some $E_6$
models{\cite {E6}}) and how these new limits compare with those already
obtained by searches using only leptonic channels{\cite {tevl}}. As we
will see, the
dijet data does yield improved bounds on new gauge bosons for certain classes
of extended gauge models particularly those in which the new $W'$ or $Z'$ is
either strongly coupled to quarks or possess a large hadronic branching
fraction. There are, of course, many extended electroweak models on the
market so that the particular set of examples we have chosen below are meant
only to be representative and not exhaustive.

Our procedure, for a given model, is to calculate the production cross section
times hadronic branching fraction($\sigma B_h$) for the new gauge
boson(or diquark) in the narrow width approximation as a
function of its mass and then perform a comparison with the limits as
quoted by CDF. To be
specific, we performed these calculations assuming a top-quark mass of 150 GeV,
$sin^2 \theta_w = 0.2325${\cite {smrev}}, and have employed the MTS1 set of
parton distribution functions of Morfin and Tung{\cite {MT}}. In
addition, we also
included a `K-factor' in the $W'$ and $Z'$ production process as given by the
work{\cite {kfact}} of Hamberg {\etal} as well as full three-loop QCD
corrections{\cite {QCD}} to the $W'$ and $Z'$
decays. To be concrete, we have also assumed that these new particles can decay
only into the conventional fermions of the SM and that any CKM-like matrices
required by the extended model
are essentially diagonal. We have also neglected any possible mixing between
the $W',Z'$ and the conventional gauge bosons of the SM as suggested by data
from LEP{\cite {smrev}}, muon decay, and radiative corrections analyses.

We will considered the new gauge bosons from the following models in
our discussion
below: ($i$) The Left-Right Symmetric Model(LRM){\cite {LRM}}, which predicts
the existence of both a $Z'$ and a $W'$. Essentially the only free parameter
in this scenario is the ratio of the right- to left-handed gauge couplings,
$\kappa = g_R/g_L$. For the $Z'$ analysis we take $\kappa =1$, while for the
$W'$ case we allow $\kappa$ to vary over its `natural' range 0.55 to 1.5.
The reason for this lower bound is that $\kappa$ is constrained by the
structure of the model to values greater than $tan \theta_w$ at which point
the couplings of the $Z'$ become singular.  We remind the reader that
the embedding of this model into a larger GUT framework, such as SO(10),
automatically leads to the prediction that $\kappa \leq 1$. Of course,
phenomenologically, such a constraint need not apply but we may generally
expect that $\kappa$ should not be very different than unity. In
comparing with searches involving final state leptons, we assume that the
leptonic decay of the $W'$ is unhindered by the existence of a large mass
for the right-handed neutrino.
($ii$)The `Alternative' Left-Right Model(ALRM){\cite {ALRM}}, which also
predicts both a $W'$ and a $Z'$ but where the $W'$ {\it {cannot}} be
produced singly
in hadronic collisions as it carries a non-zero Lepton number as well as
negative R-parity. ($iii$)The superstring-`inspired', Effect Rank-5
Models(ER5M){\cite {E6}}, predict only a $Z'$ whose couplings critically
depend on a parameter $-\pi /2 \leq \theta \leq \pi /2$. We will take
two representative values for
$\theta$ below corresponding to the so-called models $\chi (\theta =-\pi /2)$
and $\psi (\theta =0)$. As examples of somewhat more `unconventional' extended
electroweak theories, we consider ($iv$)the model of Foot and Hernandez(FH)
{\cite {FH}}, wherein an additional $Z'$ arises from the breaking of an
extended color group, such as $SU(4)_C$ or $SU(5)_C$. The $Z'$ in this scenario
is quite
strongly coupled to $q\bar q$ and simultaneously only weakly coupled to
lepton pairs. As
a second example of models of this class, we examine the predictions of ($v$)
the `Un-Unified' Model(UUM){\cite {HARV}} of
Georgi \etal  ~wherein quarks and leptons couple to the gauge fields belonging
to distinct $SU(2)$ groups.
The $W'$ and $Z'$ couplings in this model depend on a single parameter,
$s_{\phi}$, which essentially lies in the range 0.35-0.95. Unlike the case
of other scenarios, the $W'$ and $Z'$ are predicted to be highly
degenerate
in this model. As a final choice, we consider($vi$) a pseudo-model wherein
the $W'$
and $Z'$ are just heavier versions of the conventional SM gauge bosons, which
we refer to as the Sequential Standard Model(SSM), and is often used by
experimentalists as a test case of extended electroweak models.

The results of our analysis for new gauge bosons searches in the dijet channel
are shown in Figs.1a-b and 2a-c for MTS1 structure functions. (Note that our
results are not greatly modified by other structure function choices  as long
as the corresponding CDF bounds are used simultaneously.) One thing we
notice immediately is that the
constraints on $W'$'s will be stronger than those on $Z'$'s. This is to be
expected as, within a generic model, the $W'$ couplings are generally somewhat
larger than those of the corresponding $Z'$ and both particles have comparable
hadronic branching fractions. We see this already in the case
of the SM where the ratio of the $W$ to $Z$ production cross sections at the
Tevatron (with appropriate leptonic branching fractions!) is of order 10.
{}From Fig. 1a we see that, unfortunately, for most extended models, the
CDF dijet data do not further
restrict the existence of a $Z'$ beyond what was already obtained from the
dilepton channel. The results for the two examples of the ER5M are quite
representative as other values of the parameter $\theta$ lead to
quantitatively similar results. However, for the FH model, we see that
the mass region
between 260 and 600 GeV is now essentially excluded. When combined with the
previously obtained limits from single jet production{\cite {FH}} we can now
conclude that the entire mass range below 600 GeV is excluded for a $Z'$ in
this model. For the LRM  with $\kappa$=1 and SSM, the dijet data rule
out only a small mass
region, 480-560 GeV; the corresponding limits from the dilepton channel
searches for
these models are 376 and 422 GeV, respectively{\cite {tevl,rizzo}}. We note
that in the LRM case, if $\kappa$ were relatively
close to 0.55, its theoretical lower bound, stronger limits on the $Z'$ mass
may be obtainable but the results we show are not significantly altered
for, say, $\kappa$=0.7.

The situation improves somewhat for the $Z'$ in the UUM as shown in Fig.1b.
For smaller values of the parameter $s_{\phi}$, the $W'$ and $Z'$ in this
model have reasonably strong couplings to quarks so that it is this region of
the parameter space that one would hope to constrain by dijet data. Fig.1b
ignores, for the moment, that the $W'$ and $Z'$ in this model are degenerate
and displays the cross sections limits for the $Z'$ only. We will see below
what happens
in the case where $Z'$ and $W'$ results are combined for this model.
{}From the figure we
see that no constraints are found for $s_{\phi}$=0.8 or 0.9 but reasonable
bounds are
obtained for smaller $s_{\phi}$ values. With $s_{\phi}$=0.4, masses as large
as 760 GeV are excluded whereas for $s_{\phi}$=0.5 (0.6,0.7), the mass range
260-630 (370-600, 470-580) GeV is excluded.  We will compare these to the
bounds obtained from the dilepton data below.

Fig.2a shows the limits obtainable from the CDF dijet data for $W'$'s
arising from both the SSM
and LRM models. For the LRM with $\kappa$=1.5, we see that the mass region
below 620 GeV is completely excluded whereas for $\kappa$=1 (0.55) only the
mass range 260-600 (490-560) is now eliminated. The limit obtained in the SSM
is identical to that of the LRM with $\kappa$=1. The corresponding lower
limits on the $W'$ mass from leptonic channel searches{\cite {tevl,rizzo}}
are 595 (520, 411) GeV for $\kappa$=1.5 (1, 0.55) and thus we see that the
dijet data can be used successfully to extend these previously known bounds.
We remind
the reader that the bounds presented here assume a right-handed CKM matrix
which is at least approximately diagonal and that all limits, from either
the dijet or leptonic channels would
be substantially degraded if this assumption were invalid. Fig.2b shows that
the constraints obtainable on the $W'$ of the UUM are not only
generally stronger
than what can be obtained in other models but are also stronger than the UUM
$Z'$ limits themselves as discussed above. If we {\it {combine}} the
$W'$ and $Z'$ UUM production cross sections and make use of the prediction
that the $W'$ and $Z'$ are
expected to be highly degenerate in this model we arrive at Fig.2c. While
limits are still not obtained in this case for $s_{\phi}$=0.9, other choices
for this parameter do result in significant bounds. For
$s_{\phi}$=0.4 (0.5, 0.6), masses
below 830 (760, 680) GeV are completely excluded by the dijet data
while the corresponding lower
bounds from the leptonic channels are 411 (455, 493)GeV. Thus the dijet data
results in a substantial improvement in the search limits for the new gauge
bosons of this model for this particular range of the $s_{\phi}$ parameter.
For $s_{\phi}$=0.7 (0.8), the mass range 220-615 (450-580) GeV is now excluded
with the corresponding lower bounds from the leptonic searches being
522(534) GeV. For $s_{\phi}$=0.9, the leptonic data yield a limit of 503 GeV.

Thus we see the general result that for many models, particularly those that
predict the existence of a $W'$ or or a $Z'$ with rather strong couplings
to $q{\bar q}$, the Tevatron dijet data can be used to greatly supplement
limits which are obtained solely from searches for leptonic decay modes.
Generally, however, models which only predict a $Z'$ and which can be
embedded in a `conventional' GUT are not very constrained by this data.

We now turn our attention to the situation of a scalar diquark, h, whose
parton-level production cross section is given in{\cite {angel}} for the
case where we
identify h with the spin-0, Q=-1/3, color-triplet object in $E_6$
models.  There are two possible choices for the Yukawa couplings of such a
particle given the structure of the $E_6$ superpotential, W; these are usually
written as{\cite {E6}}: $\lambda_9 QQh + \lambda_{10} u^cd^ch^c$. In either
case, generation indices are suppressed and the values of the Yukawa couplings
are {\it {a priori}} unknown. Since our diquark is being produced in hadronic
interactions where u- and d-quark distributions dominate, we will assume in
our discussion below that the diquark under consideration couples primarily
only to the
quarks of the first generation. If the dominant Yukawa couplings of the
diquark were instead to the quarks
of the other generations, the resulting production cross sections would be
substantially reduced. The usual procedure in the literature{\cite {E6}} is
to scale both the unknown Yukawa couplings,
$\lambda_{9,10}$, to the strength of the electromagnetic interactions, i.e.,
$\lambda^2_{9,10}/4\pi=\alpha F_{9,10}$, and treat $F_{9,10}$ as unknowns. If
these $F$'s are of order unity or less, we then find that the diquark is a
very narrow resonance with $\Gamma_h /m_h$ of order 0.01. Because of
isospin algebra, we find that the width and production cross section of h
are both a
factor of 4 times larger when $F_9$=1 then when $F_{10}$=1(in the narrow
width approximation), so we will
specifically focus on this case. Fig.3 shows h's production cross section,
assuming $F_9$=1 and $F_{10}$=0, also assuming a constant `K-factor' of 1.33
included to account for anticipated but as yet uncalculated enhancements due to
NLL QCD. (Of course, the
hadronic branching fraction for this particle is identically unity.) Shown for
comparison is the CDF bound from the dijet data under the assumption that
$\Gamma /M$=0.01 with MTS1 structure functions. As we can see immediately,
even for electromagnetic coupling
strengths, the data puts hardly any constraint as yet on the existence of
$E_6$ diquarks.
If, however, $F_9$ were significantly larger, say, $F_9$=3, we see
that diquarks as massive as 570 GeV would be ruled
out by the CDF dijet data. Limits on scalar diquarks are thus seen to be
highly sensitive to our assumptions about the size of these unknown Yukawa
couplings and the estimated size of the effective `K-factor' in the production
process. We anticipate, however, that a several-fold increase in
integrated luminosity at the Tevatron, as is expected from the 1992-3 run,
will  be able to eliminate the possibility of diquarks, with electromagnetic
coupling strength to the first generation of quarks, up to substantially
large masses of order 600 GeV or more.

The results of our analysis can be summarized by the following observations:

($i$) $Z'$'s are generally less highly constrained by the dijet data than are
$W'$'s. In models which predict only a $Z'$ and that can be embedded in a
`conventional' GUT scenario, such as in the ER5M case, essentially no new
constraints on particle masses are obtained. Limits on the $W'$ of both the
LRM and SSM are found to be significantly improved over what is obtained by
using the dilepton data alone.

($ii$) Substantial improvement in the limits on new gauge boson masses are
obtained for both the FH and the UUM cases as these scenarios predict that
the $Z'$ and/or $W'$ is relatively strongly coupled to $q\bar q$, at least
over a reasonable range of model parameters.

($iii$) For `$E_6$-like' diquarks with electromagnetic strength Yukawa
couplings to first generation quarks, no limits are obtained. {\it {However}},
a rather modest increase in the Tevatron integrated luminosity would result
in rather significant bounds being obtained. Of course, the assumption of
somewhat stronger Yukawa couplings leads to rather stringent constraints
from existing data. To fully quantify these bounds, QCD
corrections to the diquark production process need to be performed.

Hopefully a positive signature for new gauge bosons will be found in the
near future and give us a glimpse of physics beyond the Standard Model.

\vskip.25in
\centerline{ACKNOWLEDGEMENTS}

The author would like to thank J.L. Hewett for discussions related to this
analysis. This research was supported in part by the U.S.~Department of
Energy under contract W-31-109-ENG-38.

\newpage

%
\def\MPL #1 #2 #3 {Mod.~Phys.~Lett.~{\bf#1},\ #2 (#3)}
\def\NPB #1 #2 #3 {Nucl.~Phys.~{\bf#1},\ #2 (#3)}
\def\PLB #1 #2 #3 {Phys.~Lett.~{\bf#1},\ #2 (#3)}
\def\PR #1 #2 #3 {Phys.~Rep.~{\bf#1},\ #2 (#3)}
\def\PRD #1 #2 #3 {Phys.~Rev.~{\bf#1},\ #2 (#3)}
\def\PRL #1 #2 #3 {Phys.~Rev.~Lett.~{\bf#1},\ #2 (#3)}
\def\RMP #1 #2 #3 {Rev.~Mod.~Phys.~{\bf#1},\ #2 (#3)}
\def\ZP #1 #2 #3 {Z.~Phys.~{\bf#1},\ #2 (#3)}
\def\IJMP #1 #2 #3 {Int.~J.~Mod.~Phys.~{\bf#1},\ #2 (#3)}

\newpage

%
{\bf Figure Captions}
\begin{itemize}

\item[Figure 1.]{Cross-section times hadronic branching fraction for $Z'$
production at the Tevatron. (a) From top to bottom, the monotonically falling
curves correspond to the FH, SSM, LRM, ALRM, $\psi$, and $\chi$ models
respectively while the upper(lower) wavy dotted line corresponds to the CDF
upper limits assuming $\Gamma/M$=0.05(0.01). (b) Same as (a), but for the UUM.
The top solid curve corresponds to $s_{\phi}=0.4$ with each subsequently lower
curve reflecting an increase in $s_{\phi}$ by 0.1.}
\item[Figure 2.]{Same as Fig. 1 but for $W'$ production. (a) Comparison of
the predictions for the LRM with the CDF bounds assuming the parameter
$\kappa$ takes the values
1.5(dash-dots), 1(solid), or 0.55(dashes). The SSM prediction is identical to
that for the LRM with $\kappa =1$. (b) Same as Fig. 1b but for the $W'$ of the
UUM scenario.(c) Same as (b) but assuming that the UUM $W'$ and $Z'$ are
degenerate.}
\item[Figure 3.]{Production cross section times hadronic branching fraction
for a spin-0, Q=-1/3, color triplet
diquark of the $E_6$ type with its Yukawa couplings scaled to electromagnetic
strength, assuming a K-factor of 1.33, and $F_{10}$=0, $F_9$=1 (solid) or
3 (dashes).
The CDF bound for a narrow resonance, $\Gamma/ M$=0.01, is also shown.}

\end{itemize}


\begin{thebibliography}{99}
\bibitem{smrev}
See talks by S.C.C.\ Ting, R.K.\ Ellis, and W.J.\ Marciano at the {\it
1992 Division of Particles and Fields Meeting}, Fermilab, November 1992.
\bibitem{ual}
UA1 Collaboration, C.\ Albajar \etal , \PLB B198 271 1987 ;
UA2 Collaboration, R.\ Ansari \etal , \PLB B195 613 1987 .
\bibitem{tevl}
CDF Collaboration, F. Abe \etal ,  \PRL 67 2609 1991 ~and \PRL 68 1463 1992 .
\bibitem{LEP}
For a recent analysis, see E.\ Nardi, E.\ Roulet, and D.\ Tommasini,
\PRD D46 3040 1992 .
\bibitem{lphil}
X.-G.~He, \etal, \PRD D44 2118 1991 .
\bibitem{UA2}
UA2 Collaboration, J.\ Alitti \etal , \ZP C49 17 1991 .
\bibitem{tevjj}
M.\ Incagli, CDF Collaboration, talk given at the {\it {Division of Particles
and Fields Meeting}}, Fermilab, November, 1992.
\bibitem{axi}
P.\ Frampton and S.\ Glashow, \PLB B190 157 1987 ;
F.\ Cuypers and P.\ Frampton, \PRL 60 1237 1988  ~and \PRL 63 125 1989 ;
F.\ Cuypers, \PLB B206 361 1988  ~and \PRD D41 3515 1990 ;
A.\ Falk, \PLB B230 119 1989 ; S.F.\ Noaves and A.\ Raychaudhuri,
\IJMP A5 2269 1990 ; R.\ Robinett and T.G.\ Rizzo, \PLB B215 777 1988 ;
L.\ Bergstrom, \PLB B206 361 1988 ; L.\ Sehgal and M.\ Wanninger,
\PLB B200 211 1988 ; M.A.\ Doncheski \etal , \PLB B206 361 1988 ;
E.D.\ Carlson \etal , \PLB B202 281 1988 ; J.A.\ Bagger \etal ,
\PRD D37 1188 1988 .
\bibitem{E6}
For a review of the $E_6$ models discussed here, see J.L. Hewett and T.G.
Rizzo, \PR 183 193 1989  .
\bibitem{MT}
J.C. Morfin and W.-K. Tung, \ZP C52 13  1991 .
\bibitem{kfact}
R. Hamberg, J. van Neervan, and T. Matsuura, \NPB B359 343 1991 .
\bibitem{QCD}
L.R.~Surguladze and M.A.~Samuel, \PRL 66 560 1990 , Erratum, {\it
ibid.}, {\bf 66}, 2416 (1991); S.G.~Gorishny \etal, \PLB B259 144 1991 .
\bibitem{LRM}
For a review of the LRM and original references, see R.N. Mohapatra, {\it
Unification and Supersymmetry}, (Springer, New York, 1986);
P. Langacker and S. Uma Sankar, \PRD D40 1569 1989 .
\bibitem{ALRM}
E. Ma, \PRD D36 274 1987 ~and \MPL A3 319 1988 ;
K.S. Babu \etal , \PRD D36 878 1987 ;
V. Barger and K. Whisnant, \IJMP A3 879 1988 ;
J.F. Gunion \etal , \IJMP A2 118 1987 ;
T.G. Rizzo, \PLB B206 133 1988 .
\bibitem{FH}
R.~Foot and O.~Hern\`andez, \PRD D41 2283 1990 ; R.~Foot, O.~Hern\`andez,
and T.G.~Rizzo, \PLB B246 183 1990 , and \PLB B261 153 1991 .
\bibitem{HARV}
H. Georgi, E.E. Jenkins, and E.H. Simmons, \PRL 62 2789 1989 ~and
\NPB B331 541 1990 ;
V. Barger and T.G. Rizzo, \PRD D41 946 1990 ;
T.G. Rizzo, \IJMP A7 91 1992 .
\bibitem{rizzo}
T.G.\ Rizzo, talk given at the {\it Workshop on Extended Gauge Model
Phenomenology}, Gyongyostarjan, Hungary, September 1992.
\bibitem{angel}
V.D.\ Angelopoulos \etal , \NPB B292 59 1987 .
\end{thebibliography}
\end{document}